\documentclass[singlecolumn,showpacs,preprint,amsmath,amssymb,superscriptaddress]{revtex4-1}
\usepackage{here}
\usepackage{color}
\usepackage[dvipdfmx]{graphicx}
\usepackage[hang,small,bf]{caption}
\usepackage[subrefformat=parens]{subcaption}
\draft % marks overfull lines with a black rule on the right

\begin{document}

%\preprint{}
%\Large
\title{Magnetic anisotropy of the van der Waals ferromagnet Cr$_2$Ge$_2$Te$_6$ studied by angular-dependent XMCD
% (Draft)
} %Title of paper



%\author{}
%\thanks{}
%\affiliation{}
%\large
\author{M. Suzuki}
\affiliation{Department of Physics, The University of Tokyo, Bunkyo-ku, Tokyo 113-0033, Japan}
\author{B. Gao}
\affiliation{Rutgers Center for Emergent Materials and Department of Physics and Astronomy, Rutgers University, Piscataway, New Jersey 08854, USA}
\author{G. Shibata}
\affiliation{Department of Physics, The University of Tokyo, Bunkyo-ku, Tokyo 113-0033, Japan}
\affiliation{Department of Applied Physics, Tokyo University of Science, Katsushika-ku, Tokyo 125-8585, Japan}
\author{S. Sakamoto}
\affiliation{Department of Physics, The University of Tokyo, Bunkyo-ku, Tokyo 113-0033, Japan}
\author{Y. Nonaka}
\affiliation{Department of Physics, The University of Tokyo, Bunkyo-ku, Tokyo 113-0033, Japan}
\author{K. Ikeda}
\affiliation{Department of Physics, The University of Tokyo, Bunkyo-ku, Tokyo 113-0033, Japan}
\author{Z. Chi}
\affiliation{Department of Physics, The University of Tokyo, Bunkyo-ku, Tokyo 113-0033, Japan}
\author{Y.-X. Wan}
\affiliation{Department of Physics, The University of Tokyo, Bunkyo-ku, Tokyo 113-0033, Japan}
\author{T. Takeda}
\affiliation{Department of Electrical Engineering and Information Systems, The University of Tokyo, Bunkyo-ku, Tokyo 113-8656, Japan}
\author{Y. Takeda}
\affiliation{Materials Sciences Research Center, Japan Atomic Energy Agency (JAEA), Sayo, Hyogo 679-5148, Japan}
\author{T. Koide}
\affiliation{Photon Factory, Institute of Materials Structure Science, High Energy Accelerator Research Organization, Tsukuba, Ibaraki 305-0801, Japan}
\author{A. Tanaka}
\affiliation{Department of Quantum Matter, ADSM, Hiroshima University, Higashi-hiroshima 739-8527, Japan}
\author{M. Kobayashi}
\affiliation{Department of Electrical Engineering and Information Systems, The University of Tokyo, Bunkyo-ku, Tokyo 113-8656, Japan}
\affiliation{Center for Spintronics Research Network, The University of Tokyo, Bunkyo-ku, Tokyo 113-0033, Japan}
\author{S.-W. Cheong}
\affiliation{Rutgers Center for Emergent Materials and Department of Physics and Astronomy, Rutgers University, Piscataway, New Jersey 08854, USA}
\author{A. Fujimori}
\affiliation{Department of Physics, The University of Tokyo, Bunkyo-ku, Tokyo 113-0033, Japan}
\affiliation{Materials Sciences Research Center, Japan Atomic Energy Agency (JAEA), Sayo, Hyogo 679-5148, Japan}
\affiliation{Department of Applied Physics, Waseda University, Shinjuku-ku, Tokyo 169-8555, Japan}


\date{\today}
\newpage
%\large
%\renewcommand{\baselinestretch}{1.5}
\begin{abstract}
%\large

The van der Waals ferromagnet Cr$_2$Ge$_2$Te$_6$ (CGT) has a two-dimensional crystal structure where each layer is stacked through van der Waals force. 
We have investigated the nature of the ferromagnetism and the weak perpendicular magnetic anisotropy (PMA) of CGT by means of X-ray absorption spectroscopy and X-ray magnetic circular dichroism (XMCD) studies of CGT single crystals. The XMCD spectra at the Cr $L_{2,3}$ edge for different magnetic field directions were analyzed on the basis of the cluster-model multiplet calculation. 
The Cr valence is confirmed to be 3+ and the orbital magnetic moment is found to be nearly quenched, as expected for the high-spin $t_{2g}$$^3$ configuration of the Cr$^{3+}$ ion. 
A large ($\sim 0.2$ eV) trigonal crystal-field splitting of the $t_{2g}$ level caused by the distortion of the CrTe$_6$ octahedron has been revealed, while the single-ion anisotropy (SIA) of the Cr atom is found to have a sign {\it opposite} to the observed PMA and too weak compared to the reported anisotropy energy. 
The present result suggests that anisotropic exchange coupling between the Cr atoms through the ligand Te $5p$ orbitals having strong spin-orbit coupling has to be invoked to explain the weak PMA of CGT, as in the case of the strong PMA of CrI$_3$. 
\end{abstract}

%\pacs{}% insert suggested PACS numbers in braces on next line

\maketitle %\maketitle must follow title, authors, abstract and \pacs

%\section{}
%\label{}
%\subsection{}
%\subsubsection{}

%\renewcommand{\baselinestretch}{1.5}
\section{Introduction}

The discovery of graphene has invoked renewed interest in the studies on two-dimensional (2D) materials including layered transition-metal chalcogenides \cite{graphene_first, review_2, review_3, TMD_review}.
The 2D materials show various fascinating properties that may lead to applications for new spintronics devices. 
In addition, the relation between dimensionality and various physical properties has long been a major subject from a fundamental point of view. For instance, the Mermin-Wagner theorem states that a long-range magnetic order in the 2D isotropic Heisenberg model  cannot be realized at finite temperature due to long-range magnetic fluctuations \cite{Mermin}. 
Recently, van der Waals (vdW) ferromagnets, which have 2D crystal structures but show ferromagnetic order \cite{vdWFerro_review}, such as Cr$_2$Ge$_2$Te$_6$ (CGT) \cite{CGT_oldcalc_prop, CGT_oldcalc_2, CGT_tricritical_exp, CGT_aniso_exp, CGT_MH_MT, CGT_bilayer, CGT_hydropressure, CGT_thinfilm_mag, ARPES1, ARPES2, EM_19_Kang} and CrI$_3$ \cite{CrI3_mono, CrI3_giant} have attracted great interest.
Although the magnetism of a few to several-monolayer CGT shows soft behaviors well described by the 2D Heisenberg model \cite{CGT_bilayer}, the finite Curie temperature of 61 K \cite{CGT_aniso_exp, CGT_oldcalc_prop, CGT_tricritical_exp} and 
the easy magnetization axis along the $c$-axis (out-of-plane) direction \cite{CGT_aniso_exp, CGT_oldcalc_prop, CGT_MH_MT} indicate a finite perpendicular magnetic anisotropy (PMA) and hence a departure from the 2D Heisenberg model. 
On the other hand, CrI$_3$ exhibits ferromagnetism down to monolayer thickness, which suggests a stronger PMA than CGT and is well described by the 2D Ising model \cite{CrI3_mono}. 

The magnetic anisotropy of a 2D ferromagnet arises primarily from (i) the single-ion anisotropy (SIA) of the magnetic ion and (ii)  anisotropic exchange interaction between the magnetic ions in the 2D layer. 
The SIA of the Cr-based vdW ferromagnets is considered to be small because of the $t_{2g\uparrow}$$^3$ configuration of the Cr$^{3+}$ ion, which has no orbital degree of freedom, as discussed, e.g., by Lado and Fern\'{a}ndez-Rossi \cite{CrI3_aniso_calc}. 
On the contrary, first-principles estimates of exchange coupling by Xu {\it et al.} \cite{CrI3 Kitaev vs SIA} have shown that the SIA and Kitaev-type anisotropic magnetic coupling in the Cr-based vdW ferromagnets are of comparable magnitudes and compete with each other. 
Furthermore, Kim {\it et al.} \cite{PRL_19_Kim} have performed multi-site cluster-model multiplet calculation and attributed the strong PMA of CrI$_3$ to anisotropic exchange coupling between Cr atoms through the iodine $5p$ ligand orbitals, which has strong spin-orbit coupling (SOC).  
In fact, X-ray magnetic circular dichroism (XMCD) measurements at the iodine $L_1$ edge of CrI$_3$ have revealed a finite orbital magnetic moment of the I $5p$ electrons, implying the importance of the ligand $p$ orbitals in the PMA of the vdW ferromagnets \cite{Choi I XMCD}. 
As for CGT, which shows a weaker PMA than CrI$_3$, the situation is complicated and it is far from clear which of the SIA or the anisotropic exchange is responsible for the PMA \cite{PRL_19_Kim, PCCP_19_Wang}.

In order to evaluate the contribution of the SIA of Cr to the magnetic anisotropy, XMCD studies at the Cr $L_{2,3}$ edge are expected to give essential information. 
So far Cr $L_{2,3}$-edge XMCD has been reported for CGT \cite{Fumega XMCD} and CrI$_3$ \cite{Frisk_XMCD} with line-shape analysis using cluster-model calculation, 
but no information about the contributions of the Cr $3d$ electrons to the PMA has been explored. 
In the present work, we have carried out Cr $L_{2,3}$-edge XMCD measurements of CGT single crystal under magnetic fields applied to different crystallographic directions and subsequent cluster-model multiplet calculation to estimate the SIA energy of the Cr atom centered in the CrTe$_6$ octahedron. 

\section{methods}

Single crystals of CGT were synthesized using a flux method. High purity Cr, Ge and Te powders were mixed in a molar ratio of 2:6:36; the extra Ge and Te were used as a flux. The mixture was loaded in an alumina crucible and sealed in an evacuated quartz tube. A ball of alumina fiber was placed on top of the crucible. The quartz tube was placed in a furnace, heated to 700 $^{\circ}$C, held for 5 hours, and then slowly cooled down to 480 $^{\circ}$C over a period of 3 days. This was followed by centrifugation to remove the flux.

X-ray absorption spectroscopy (XAS) and XMCD measurements at the Cr $L_{2,3}$ edge were performed at beam line BL23SU of SPring-8 and beam line BL-16A of Photon Factory at High Energy Accelerator Research Organization (KEK-PF).
The measurement at SPring-8 was performed under a magnetic field of $\mu_0H= 3$ T parallel to the $c$-axis of the sample. The XAS and XMCD spectra were recorded in the total electron yield mode. Right- ($\sigma^+$) and left ($\sigma^-$) circularly polarized x-rays were switched at a frequency of 1 Hz in order to eliminate time-dependent background from the obtained XMCD spectra. 
The sample temperature ($T$) during the measurement was set to 20 K, well below the Curie temperature. The sample was cleaved in the vacuum chamber prior to the measurements to obtain fresh surfaces. Magnetic-field-direction dependence of XMCD spectra was measured at the BL-16A of KEK-PF. Polarization switching was made at a frequency of 10 Hz. Magnetic field of  $\mu_0H = 1$ T was applied parallel and perpendicular to the $c$-axis. Signals were recorded in the total electron-yield mode.

\begin{figure}[H]
  \begin{center}
   \includegraphics[width=12cm]{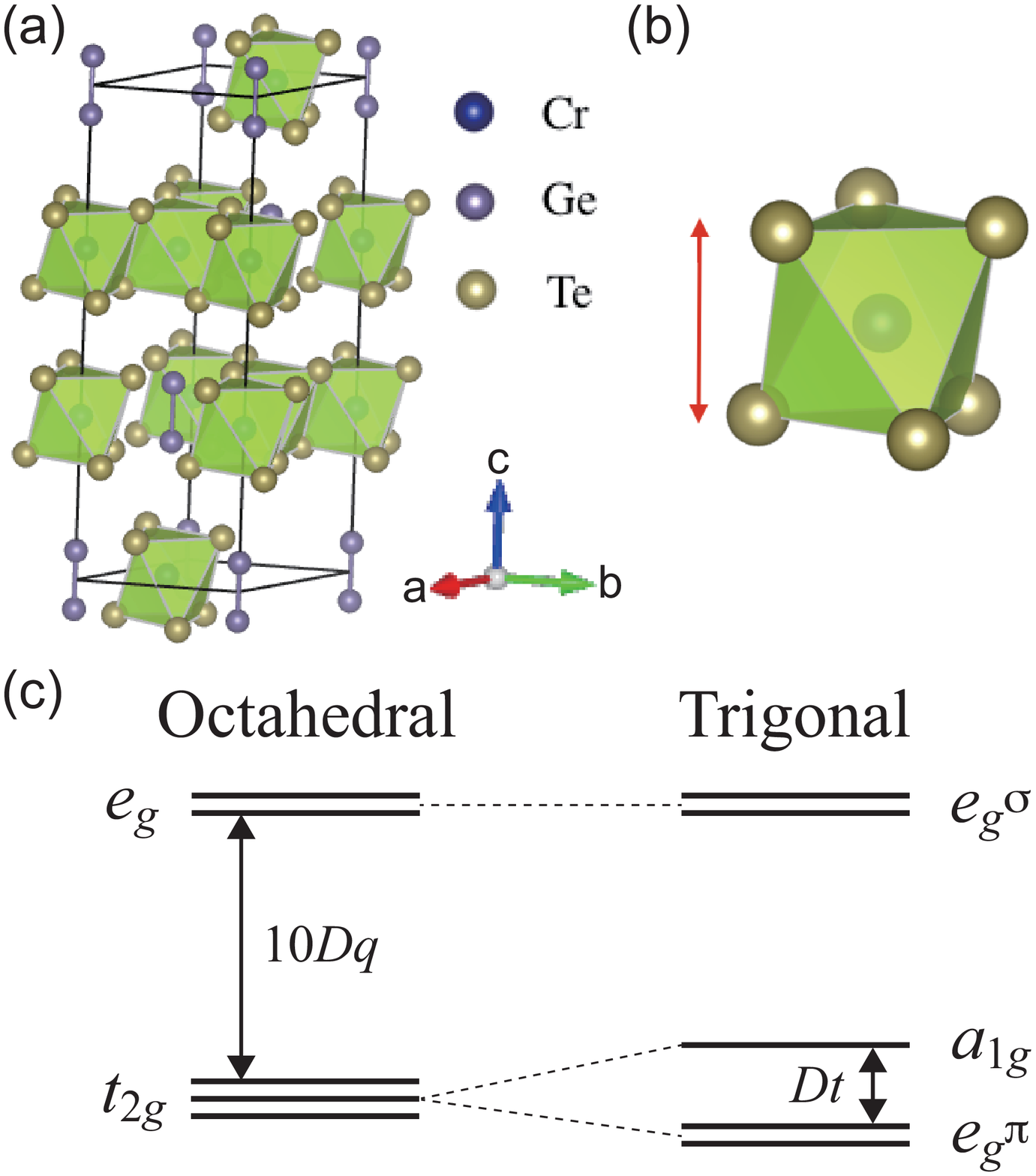}
  \end{center}
  \label{energy_levels}
  \caption{Crystal structure of Cr$_2$Ge$_2$Te$_6$ (CGT)
(a) Schematic drawing of the crystal structure. (b) CrTe$_6$ octahedron which is elongated along the $c$-axis. (c) Crystal-field splitting of the Cr 3$d$ orbital in the distorted octahedron to trigonal symmetry.}
  \label{CrTe6_cluster}
\end{figure}

To investigate the electronic structure which contributes to the ferromagnetism, configuration-interaction cluster-model multiplet calculations have been conducted.
Figure \ref{CrTe6_cluster}(a) shows the crystal structure of CGT.  
Each Cr atom is surrounded by six Te atoms, forming a CrTe$_6$ cluster.  
It constitutes the CrGeTe$_{3}$ layer, which are stacked through vdW force. 
The CrTe$_6$ cluster is slightly elongated along the $c$-axis \cite{CGT_oldcalc_prop}, as shown in Fig. \ref{CrTe6_cluster}(b), 
which means that the Cr electrons are under a nearly octahedral, trigonally distorted crystal field. 
The following parameters of the cluster model are used to reproduce the experimental XMCD spectra: the Cr 3$d$-3$d$ Coulomb energy $U_{dd}$, the Cr 2$p$-3$d$ Coulomb energy $U_{pd} \approx 1.2U_{dd}$, the charge-transfer energy $\Delta$ from the Te 5$p$ to Cr 3$d$ levels, the Slater-Koster parameter ($pd\sigma$) between the Cr 3$d$ and Te 5$p$ orbitals, the crystal-field splitting 10$Dq$ ($\approx 0.7$ eV) between the Cr $e_{g}$ and $t_{2g}$ levels (apart from the $p$-$d$ covalency contribution), and the trigonal crystal-field splitting $Dt$ between the $a_{1g}$ and $e^{\pi}_{g}$ orbitals. [For the definition of 10$Dq$ and $Dt$, see Fig.~\ref{CrTe6_cluster}(c).]
As for the parameter values, we follow the empirical relationship  \cite{fujimori_para} and employ $\Delta = 4.5$ eV, $U_{dd}=6.0$ eV, and $(pd\sigma)=2.6$ eV as being relevant to Cr$^{3+}$ tellurides. The SOC parameter of the Cr $3d$ orbital is taken to be 35 meV.

\section{results and discussion}

\begin{figure}[H]
\begin{center}
\includegraphics[width = 12cm]{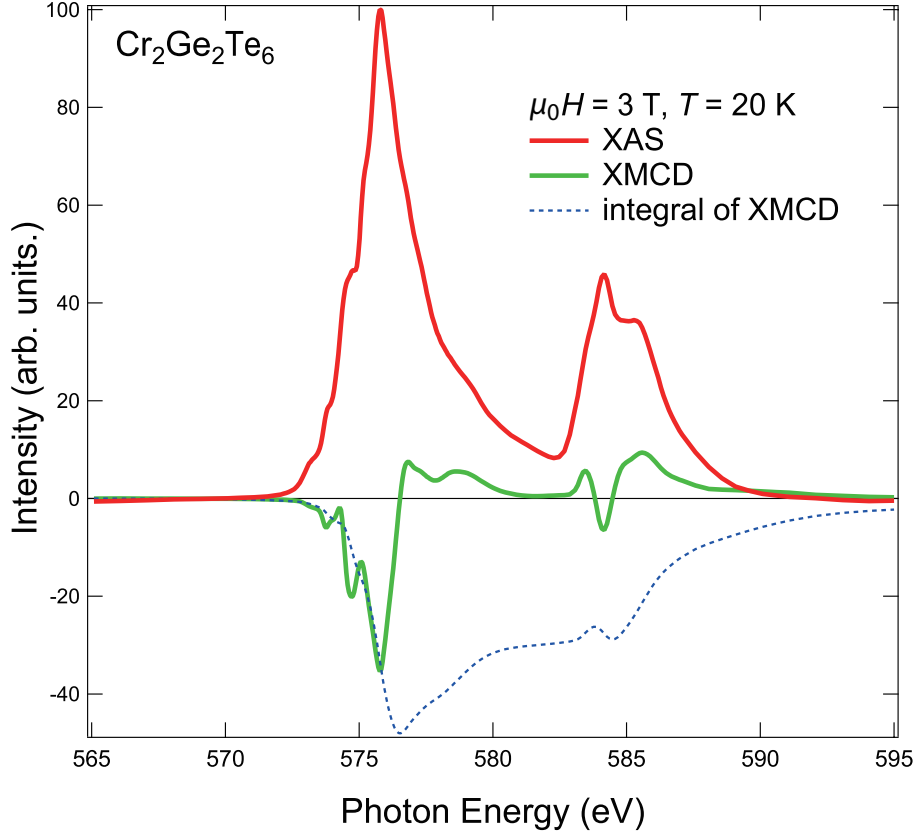}
\end{center}
\caption{
Cr $L_{2,3}$ XAS and XMCD spectra of CGT at $\mu_0H$ = 3 T and $T$ = 20 K. The red curve shows the average of XAS spectra obtained with left- and right-circularly polarized x-rays. The green curve shows the XMCD spectrum, which is the intensity difference between the above two XAS spectra. The blue dotted curve shows the integral of the XMCD spectrum. Magnetic field is applied perpendicular to the sample surface ($H \parallel c$-axis), that is, along the easy magnetization axis.}
\label{XAS_XMCD}
\end{figure}

Figure \ref{XAS_XMCD} shows Cr $L_{2,3}$-edge XAS and XMCD spectra of CGT under the out-of-plane magnetic field of $\mu_0H =$ 3 T at $T$ = 20 K. The applied magnetic field was high enough to saturate the magnetization of CGT because the magnetization saturates below $\sim 0.5$ T \cite{CGT_tricritical_exp}.
The tail of the Te $M_{4,5}$ XAS peak overlapping the Cr $L_{2,3}$ XAS spectrum has been subtracted assuming a hyperbola. Arctangent steps at the $L_2$ and $L_3$ edges have also been subtracted from the Cr $L_{2,3}$ XAS as usual. 
On the other hand, the XMCD signals predominantly come from the magnetic Cr atoms, and there is no need to subtract signals from the Te atoms. 
 
\begin{figure}[H]
\begin{center}
\includegraphics[width = 12cm]{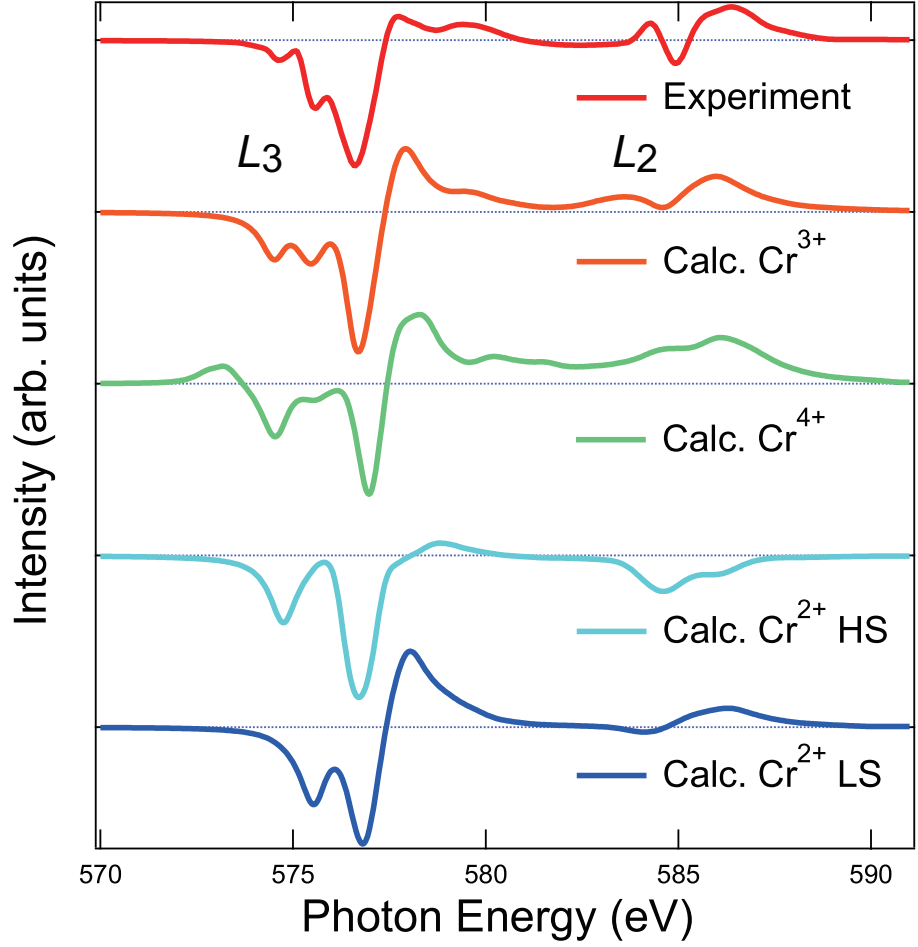}
\end{center}
%\vspace{0.5cm}
\caption{
Experimental and calculated Cr $L_{2,3}$ XMCD spectra of CGT. The experimental spectrum is the XMCD spectrum taken with magnetic field perpendicular to the $ab$ plane. The orange, green, light-blue, and blue curves show XMCD spectra calculated using the cluster model, assuming that the valence of Cr is 3+, 4+, 2+ high-spin (HS), and 2+ low-spin (LS) states, respectively. }
\label{XMCD_23}
\end{figure}

Figure \ref{XMCD_23} shows comparison of the measured Cr $L_{2,3}$ XMCD spectrum with those calculated using the cluster model assuming the Cr valences of 4+, 3+, 2+ high-spin states, and 2+ low-spin states. 
The experimental spectrum shows three negative peaks and one positive peak at the $L_3$ edge, and one negative peak and two positive peaks at the $L_2$ edge. 
The comparison shows that the calculation for the Cr$^{3+}$ state reproduces the experimental spectrum better than those for the Cr$^{2+}$ and Cr$^{4+}$ states. 
Thus, one can conclude that the Cr atom in CGT is in the $3+$ state (with the high-spin $t_{2g}^3$ configuration), and not in the $2+$ nor $4+$ state. 

\subsection{Spin and orbital magnetic moments}

The orbital and effective spin magnetic moments are quantitatively estimated from the XAS and XMCD spectra using the XMCD sum rules \cite{Carra, Thole}.
For the $L_{2,3}$ edge ($2p \rm \rightarrow 3d$ transition), the sum rules are given by:
\begin{eqnarray*}
m_{\rm orb} &=& - \frac{4n_{\rm h}} {3}\frac{\int_{L_3 + L_2}(\mu_+ - \mu_-)d\omega}{\int_{L_3 + L_2}(\mu_+ + \mu_-)d\omega},\\
m_{\rm spin}^{\rm eff} &=& m_{\rm spin} + 7m_{\rm T} \\ &=& -{2n_{\rm h}} \frac{\int_{L_3}(\mu_+ - \mu_-)d\omega - 2\int_{L_2}(\mu_+ - \mu_-)d\omega}{\int_{L_3 + L_2}(\mu_+ + \mu_-)d\omega} ,
\end{eqnarray*}
where $\mu _ +$ ($\mu _ -$) is an XAS spectrum taken with $\sigma^+$ ($\sigma^-$), $m_{\rm orb}$, $m_{\rm spin}$ and $m_{\rm T}$ are the orbital magnetic moment, spin magnetic moment, and magnetic dipole moment, respectively, in units of Bohr magneton $\mu_{\rm B}$/Cr atom. 
$\mu _ +$ ($\mu _ -$) stands for the absorption coefficient for photon helicity parallel (antiparallel) to the Cr majority spin direction. 
$n_h$ is the number of holes in the valence shell. 
By applying the sum rules to the spectra in Fig. \ref{XAS_XMCD}, the orbital magnetic moment of $m_{\rm orb} = 0.00 \pm 0.05$  $\mu_{\rm B}$/Cr atoms has been obtained, consistent with the quenched orbital magnetic moment in the $d^3$ configuration of Cr$^{3+}$. 
The effective spin magnetic moment of  $m_{\rm spin}^{\rm eff} \equiv m_{\rm spin} + 7m_{\rm T} = 2.2 \pm 0.1$ $\mu_{\rm B}$/Cr atom has been obtained, where the $m_{\rm spin}$ and $m_{\rm T}$ are the spin magnetic moment and the magnetic dipole moment, respectively. 
In a high-symmetry environment of the Cr atom as in CGT, $m_{\rm T}$ is negligibly small compared to $m_{\rm spin}$ \cite{dipole_1, dipole_2}. 
Therefore, since $m_{\rm orb}$ is nearly zero, we regard $m_{\rm spin}^{\rm eff}$ as the total magnetic moment $m_{\rm spin}$. 
The value $m_{\rm spin} \sim m_{\rm spin}^{\rm eff} = 2.2 \pm 0.1 \mu_{\rm B}$ is consistent with that of the reported saturation magnetization of CGT, 2.2 - 2.9 $\mu_{\rm B}$/Cr atom \cite{CGT_MH_MT, CGT_aniso_exp, CGT_oldcalc_prop}. 

\subsection{Crystal-field splitting}

\begin{figure}[H]
\begin{center}
\includegraphics[width = 12cm]{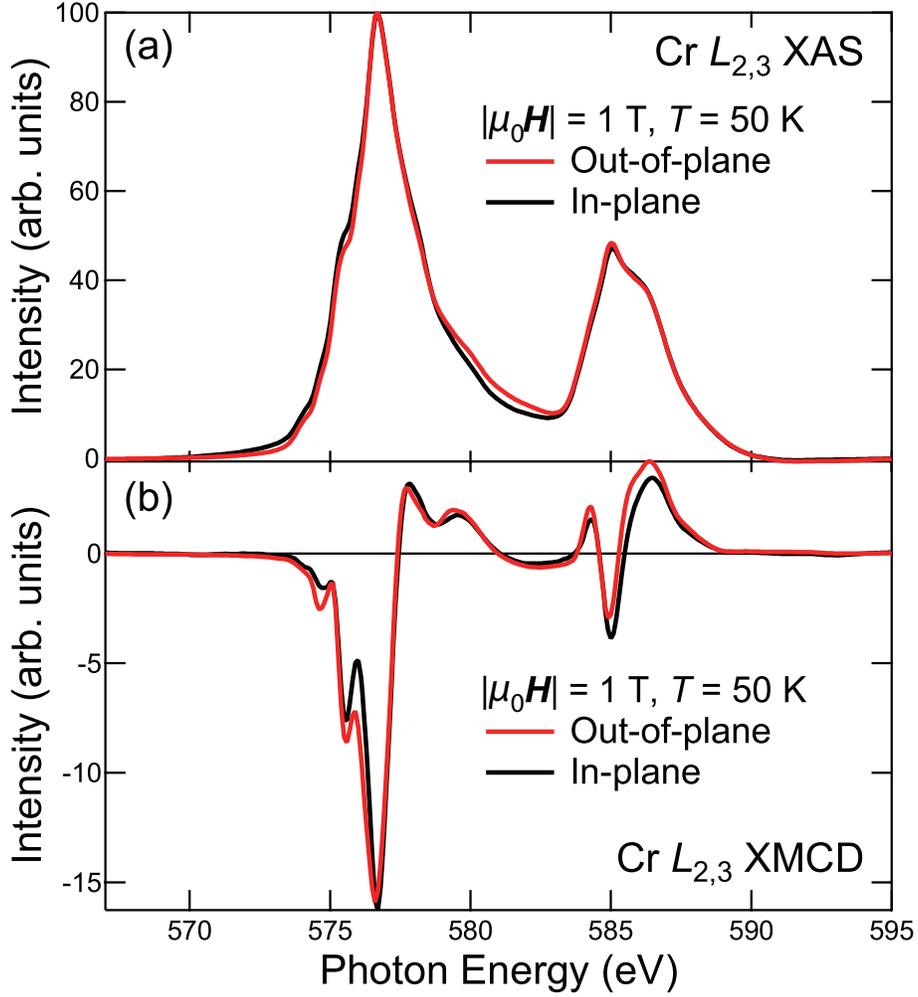}
\end{center}
%\vspace{1.5cm}
\caption{Experimental Cr $L_{2,3}$ XAS and XMCD spectra of CGT obtained with different magnetic field directions. (a) Red and black curves show XAS spectra taken with the out-of-plain and in-plane magnetic fields, respectively. (b) XMCD spectra under out-of-plane and in-plane magnetic fields. Red and black spectra show the out-of-plane and in-plane XMCD spectra, respectively. The strength of the magnetic field was 1 T and the measurement temperature was 50 K.}
\label{XAS_XMCD_dir}
\end{figure}

\begin{figure}[H]
\begin{center}
\includegraphics[width = 12cm]{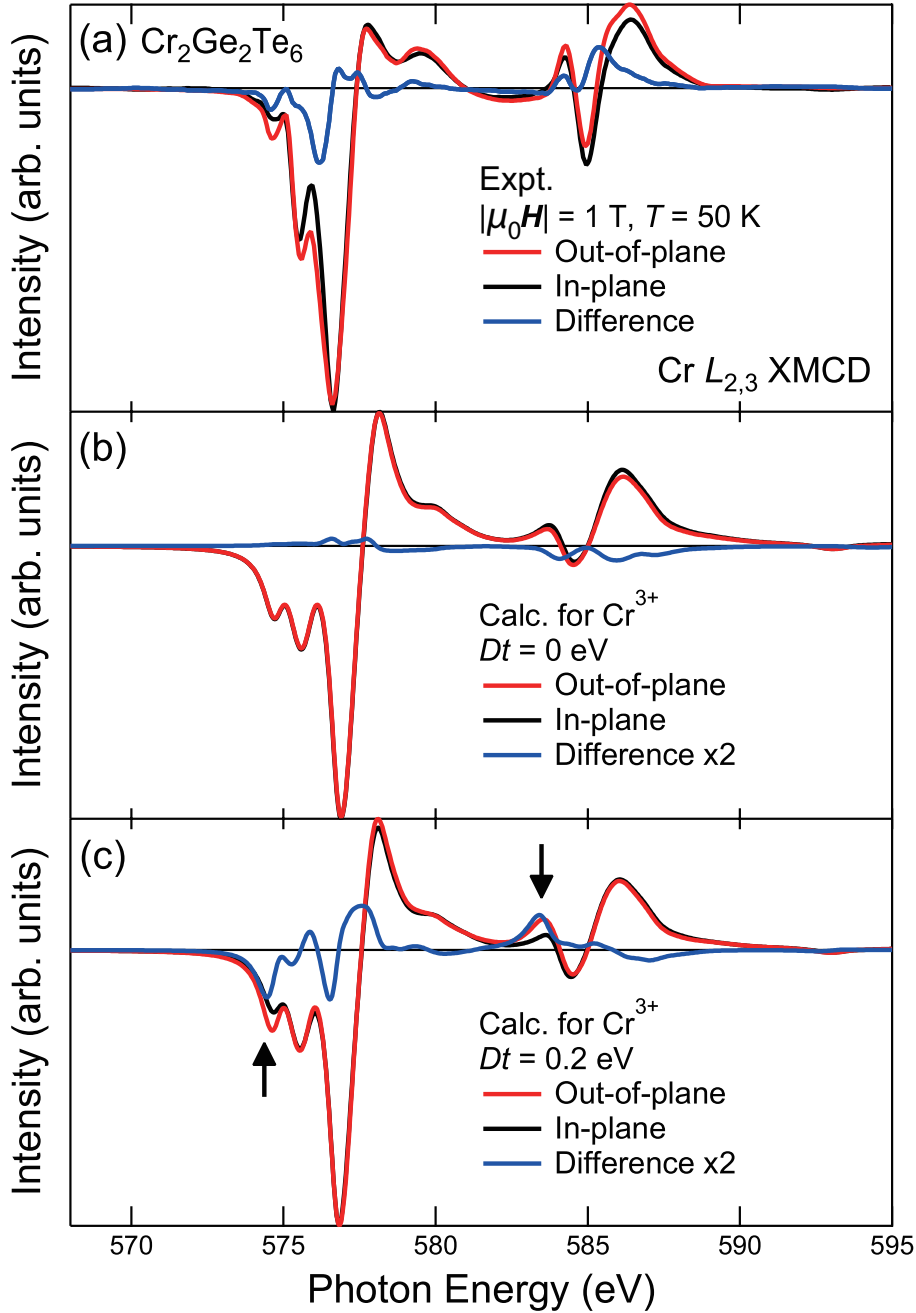}
\end{center}
%\vspace{0.5cm}
\caption{ Experimental and calculated Cr $L_{2,3}$ XMCD spectra of CGT for different magnetic field directions. The red and black curves show XMCD spectra taken under magnetic fields perpendicular (out-of-plane) and parallel (in-plane) to the surface, respectively. Blue curves show the difference between the XMCD spectra measured with different magnetic field directions. (a) Experimental XMCD spectra. 
(b), (c) XMCD spectra obtained by cluster-model calculation with and without the trigonal crystal field. $Dt$ denotes the trigonal-field splitting. Clear differences in the line shapes the different magnetic field directions are indicated by black arrows.}
\label{XMCD_compare}
\end{figure}

In order to study the anisotropies of the electronic and magnetic properties of CGT, XMCD measurements under different magnetic field directions have been made.
Figure \ref{XAS_XMCD_dir} shows the Cr $L_{2,3}$ XAS and XMCD spectra taken with magnetic fields applied to the out-of-plane ($H \parallel c$-axis) and in-plane ($H \parallel ab$-plane) directions. 
The XMCD spectra show small but clear differences between the different magnetic-field directions, while the spectral line shapes of the XAS spectra are relatively insensitive to the field direction. 
To identify the origin of the spectral differences between the different field directions, cluster-model multiplet calculations have been conducted assuming the Cr valence to be 3+. 
The effects of the elongation of the CrTe$_6$ octahedron along the $c$-axis \cite{CGT_oldcalc_prop} are taken into account as a trigonal crystal-field splitting $Dt$ of the $t_{2g}$ level as shown in Fig. \ref{CrTe6_cluster}(c),
where the triply-degenerate $t_{2g}$ level is split into the non-degenerate $a_{1g}$ and doubly-degenerate $e^{\pi}_{g}$ levels. 
Here, for a positive (negative) trigonal crystal field $Dt$, the $a_{1g}$ level is located above (below) the $e^{\pi}_{g}$ level.
Figures \ref{XMCD_compare}(b) and (c) show the calculated spectra without and with the trigonal crystal field, respectively, compared with experiment [Fig.~\ref{XMCD_compare}(a)].
Without the trigonal crystal field, i.e., $Dt = 0$ [Fig.~\ref{XMCD_compare}(b)], there is no difference in the XMCD spectra between the out-of-plain and in-plane magnetic fields. 
On the other hand, the calculated XMCD spectra under the trigonal crystal field $Dt$ of 0.2 eV show finite differences between the out-of-plain and in-plane magnetic fields, as shown in Fig.~\ref{XMCD_compare}(c), and well explain the experiment [Fig.~\ref{XMCD_compare}(a)]. 
Thus we conclude that the differences in the XMCD spectra taken with the different magnetic-field directions are attributed to the elongation of the CrTe$_6$ octahedron along the $c$-axis.

\subsection {Magnetic anisotropy energies}

To evaluate the SIA of the Cr ion, the ground-state energy of the cluster is calculated using the cluster model for the different magnetic field directions. 
Thus we find that the energy for the in-plane magnetic field is lower than the energy for the out-of-plane magnetic field by $\sim 5 \times 10^{-6}$ eV/Cr atom, that is, the Cr atom has the easy magnetization axis in the $ab$-plane. 
This is opposite to the observed PMA of CGT and furthermore its magnitude is 1-2 orders of magnitude smaller than the experimental value of $\sim 1-4\times10^{-5}$ eV/Cr atom \cite{CGT_aniso_exp, CGT_bilayer} as well as the first-principles-theoretical value of SIA, $\sim 3 \times 10^{-4}$ eV/Cr atom \cite{EM_19_Kang}. 
The present result, on the other hand, agrees with the first-principles calculation by Wang {\it et al.} \cite{PCCP_19_Wang} in that the SIA of Cr is weak and acts against PMA. 

In spite of the large trigonal crystal field $Dt \sim 0.2$ eV and the resulting anisotropy in the XMCD spectra, the calculated magnetic anisotropy energy is small because of the high-spin $t_{2g}^3$ configuration, which has no orbital degree of freedom.  
Therefore, we consider that effects beyond the SIA evaluated using the cluster model, where only the spin-orbit coupling of the Co $3d$ electrons are taken into account, may play important roles in realizing the PMA of CGT.  
Such effects would include anisotropic exchange coupling between Cr ions through the ligand Te $5p$ electrons with strong SOC, in analogy to CrI$_3$. 
In the case of CrI$_3$, which has a stronger PMA than CGT, density functional theory (DFT) calculation \cite{CrI3_structure_info, CrI3_aniso_calc} and multi-Cr site cluster-model multiplet calculation  \cite{PRL_19_Kim} have clearly shown that exchange coupling between Cr ions through I $5p$ orbitals explains the stronger PMA of CrI$_3$. 

In addition to the SIA of Cr and anisotropic exchange between Cr discussed above, the magnetic anisotropy of the Te $5p$ electrons themselves may contribute to the magnetic anisotropy of CGT \cite{PCCP_19_Wang}. Such an anisotropy could arise because the Te $5p$ electrons are partially spin-polarized due to hybridization with the Cr $3d$ electrons, in analogy to the magnetic anisotropy of the Pt $5d$ electrons in $L1_0$-type FePt \cite{FePt_Solovyev, FePt_Ikeda}. In order to address this issue, further theoretical and experimental (XMCD at the Te absorption edges) studies would be necessary.    

\section{conclusion}

We have performed XAS and XMCD measurements at the Cr $L_{2,3}$ edge and subsequent cluster-model calculation on CGT to investigate the electronic structure which contributes to the ferromagnetism and magnetic anisotropy. 
The effective spin magnetic moment obtained using the sum rules is 2.2 $\rm \mu_B$, consistent with previous reports, while the orbital magnetic moment is nearly zero. 
Subtle changes of the XMCD spectra are observed by changing the magnetic field direction, and are attributed to the distortion of the CrTe$_6$ octahedron from comparison with cluster-model calculation including the trigonal crystal field of the cluster elongated along the $c$ axis. 
The magnetic anisotropy energy calculated using the cluster model is too small and has an opposite sign compared to experiment.  
This implies that the magnetic anisotropy of CGT cannot be explained by the SIA of Cr described by the cluster model, and that one has to take into account interaction between Cr atoms through Te $5p$ ligands which have strong SOC.

\begin{acknowledgments}

We would like to thank I. Solovyev, D. Khomskii, S. Streltsov, and K. Kugel for enlightening discussion, 
and M. Suzuki-Sakamaki and K. Amemiya for valuable technical support at KEK-PF. 
This work was supported by a Grant-in-Aid for Scientific Research from JSPS (grant Nos. 15H02109 and and 19K03741), by Center for Spintronics Research Network (CSRN), the University of Tokyo, under Spintronics Research Network of Japan (Spin-RNJ), and by the Shared Use Program of Japan Atomic Energy Agency (JAEA) Facilities (Proposal No. 2018A-E24) supported by JAEA Advanced Characterization Nanotechnology Platform as a program of "Nanotechnology Platform" of MEXT (Proposal No. A-18-AE-0019). 
The synchrotron experiment at KEK-PF was done under the approval of the Program Advisory Committee (Proposal Nos. 2016S2-005), and the experiment at SPring-8 was done at the JAEA beam line BL23SU (Proposal Nos. 2018A3841). 
The work at Rutgers University was funded by the Gordon and Betty Moore Foundation's EPiQS Initiative through Grant GBMF4413 to the Rutgers Center for Emergent Materials. 
\end{acknowledgments}

\newpage

%\begin{figure}[htbp]

\end{document}